\documentclass[12pt]{article}
\usepackage{epsfig}
\usepackage{amssymb,amsmath}

\newcommand{\bea}{\begin{eqnarray}}
\newcommand{\eea}{\end{eqnarray}}

 \makeatletter
\def\alt{\mathrel{\mathpalette\gl@align<}}
\def\agt{\mathrel{\mathpalette\gl@align>}}
\def\gl@align#1#2{\lower.6ex\vbox{\baselineskip\z@skip\lineskip\z@
\ialign{$\m@th#1\hfil##\hfil$\crcr#2\crcr\sim\crcr}}} \makeatother

\begin{document}
\begin{flushright}
YGHP16-03
\end{flushright}
\vspace*{1.0cm}

\begin{center}
\baselineskip 20pt 
{\Large\bf 
$Z^\prime_{BL}$ portal dark matter and LHC Run-2 results 
}
\vspace{1cm}

{\large 
Nobuchika Okada$^{~a}$  and  Satomi Okada$^{~b}$
}
\vspace{.5cm}

{\baselineskip 20pt \it
$^a$Department of Physics and Astronomy, University of Alabama, Tuscaloosa, AL35487, USA\\
$^b$Graduate School of Science and Engineering, Yamagata University,  \\
Yamagata 990-8560, Japan} 

\vspace{.5cm}

\vspace{1.5cm} {\bf Abstract}
\end{center}

We consider a concise dark matter scenario in the minimal gauged $B-L$ extension of the Standard Model (SM),  
  where the global $B-L$ (baryon number minus lepton number) symmetry in the SM is gauged, 
  and three generations of right-handed neutrinos and a $B-L$ Higgs field are introduced.  
Associated with the $B-L$ gauge symmetry breaking by a VEV of the $B-L$ Higgs field, 
  the seesaw mechanism for generating the neutrino mass is automatically implemented 
  after the electroweak symmetry breaking in the SM. 
In this model context, we introduce a $Z_2$-parity and assign an odd parity 
  for one right-handed neutrino while even parities for the other fields. 
Therefore, the dark matter candidate is identified as the right-handed Majorana neutrino with odd $Z_2$ parity, 
  keeping the minimality of the particle content intact. 
When the dark matter particle communicates with the SM particles 
  mainly through the $B-L$ gauge boson ($Z^\prime_{BL}$ boson), 
  its relic abundance is determined by only three free parameters, the $B-L$ gauge coupling ($\alpha_{BL}$), 
  the $Z^\prime_{BL}$ boson mass ($m_{Z^\prime}$) and the dark matter mass ($m_{DM}$). 
With the cosmological upper bound on the dark matter relic abundance 
   we find a lower bound on $\alpha_{BL}$ as a function of $m_{Z^\prime}$. 
On the other hand, we interpret the recent LHC Run-2 results on search for $Z^\prime$ boson resonance 
   to an upper bound on $\alpha_{BL}$ as a function of $m_{Z^\prime}$. 
Combining the two results we identify an allowed parameter region for this ``$Z^\prime_{BL}$ portal'' dark matter scenario, 
   which turns out to be a narrow window with the lower mass bound of $m_{Z^\prime} > 2.5$ TeV. 

\thispagestyle{empty}

\newpage

\addtocounter{page}{-1}

\baselineskip 18pt

\section{Introduction} 
The neutrino mass matrix and the candidate of the dark matter particle 
  are major missing pieces in the Standard Model (SM), 
  that must be supplemented by the framework of beyond the SM.  
The seesaw mechanism~\cite{seesaw} is probably the most natural way 
  to incorporate the tiny neutrino masses and their flavor mixing, 
  where right-handed neutrinos with Majorana masses are introduced.  
The minimal $B-L$ extended SM~\cite{MBL} is a very simple extension of the SM 
  to naturally incorporate the seesaw mechanism. 
In this model, the accidental global $B-L$ (baryon number minus lepton number) symmetry 
  in the SM is gauged, and an introduction of the three generations of right-handed neutrinos 
  is required to keep the model from the gauge and gravitational anomalies. 
Associated with the $B-L$ gauge symmetry breaking, the right-handed neutrinos acquire 
  Majorana masses, and the light neutrino Majorana masses are generated 
  through the seesaw mechanism after the electroweak symmetry breaking. 
The mass spectrum of new particles introduced in the minimal $B-L$ model, 
  the $B-L$ gauge boson ($Z^\prime_{BL}$ boson), the right-handed Majorana neutrinos 
  and the $B-L$ Higgs boson, is controlled by the $B-L$ symmetry breaking scale. 
If the breaking scale lies around the TeV scale, the $B-L$ model can be discovered  
  at the Large Hadron Collider (LHC) in the near future.

One of the most promising candidates for the dark matter in the present universe 
   is the Weakly Interacting Massive Particle (WIMP),  which was in thermal equilibrium 
   in the early universe and its relic density is determined by the interactions with the SM particles. 
It is a prime open question in particle physics and cosmology to identify the properties of the dark matter particle. 
There are, in general, various ways to supplement a dark matter particle to the SM.   
A simple and concise way to introduce a dark matter candidate in the context of the minimal $B-L$ model 
   has been proposed in \cite{OS_DM}, where only a $Z_2$ parity is introduced without any extensions    
   of the particle content of the model. 
An odd parity is assigned to one right-handed neutrino, while the other particles have even parties. 
As a result, the $Z_2$-odd right-handed neutrino plays a role of dark matter.  
The neutrino oscillation data can be reproduced by the so-called minimal seesaw~\cite{Minimal-Seesaw} 
  with two generations of the right-handed neutrinos, predicting one massless neutrino.  
Dark matter phenomenology in this model context has been investigated~\cite{OS_DM, OO_DM}. 
The right-handed neutrino dark matter can annihilate into the SM particles through its interactions 
   with the $Z^\prime_{BL}$ boson and two Higgs bosons which are realized as linear combinations 
   of the SM Higgs and the $B-L$ Higgs bosons.  
Supersymmetric (SUSY) extension of the model has also been proposed~\cite{MSUSYBL}, 
   where the $B-L$ gauge symmetry is radiatively broken at the TeV scale 
   through SUSY breaking effects~\cite{RBLSB1, RBLSB2, MSUSYBL}.  
In the SUSY extension of the model, the right-handed neutrino dark matter communicates 
  with the SM particles only through the $Z^\prime_{BL}$ boson, 
  because SUSY forbids a mixing term between the SM Higgs and the $B-L$ Higgs fields 
  in the superpotential at the renormalizable level.

Recently, the so-called $Z^\prime$ portal dark matter has a lot of attention~\cite{Zp-protal1}-\cite{Zp-protal5}, 
   where a dark matter particle is introduced along with an extra gauge extension of the SM, 
   and the dark matter particle communicates with the SM particles through the $Z^\prime$ gauge boson. 
The $Z^\prime$ boson as a mediator allows us to investigate a variety of dark matter physics, 
  such as the dark matter relic abundance and the direct and indirect dark matter search.  
Interestingly, the search for $Z^\prime$ boson resonance at the LHC provides information 
  that is complementary to dark matter physics. 
The minimal (SUSY) $B-L$ model with the right-handed neutrino dark matter discussed above 
  is a very simple example of  the $Z^\prime$ portal dark matter model, that we investigate in this paper\footnote{
As mentioned above, the right-handed neutrino dark matter can communicate with the SM particles 
  also through the Higgs bosons and hence, it is a candidate of the so-called Higgs portal dark matter. 
See  \cite{OS_DM, OO_DM} for detailed analysis on the $Z_2$-odd right-handed  neutrino 
  as the Higgs portal dark matter. 
}. 
Because of the simplicity of the model, dark matter physics is controlled 
  by only three free parameters, the $B-L$ gauge coupling ($\alpha_{BL}$), 
  the $Z^\prime_{BL}$ boson mass ($m_{Z^\prime}$) and the dark matter mass ($m_{DM}$).  
We will identify allowed parameter regions of the model by considering the cosmological bound 
  on the dark matter relic abundance and the most recent results by the LHC Run-2 
  on search for $Z^\prime$ boson resonance with dilepton final states~\cite{ATLAS13TeV, CMS13TeV}.

This paper is organized as follows. 
In the next section, we define the minimal $B-L$ model with the right-handed neutrino dark matter.  
In Sec.~3, we analyze the relic abundance of the $Z^\prime_{BL}$ portal dark matter, 
  and identify a model parameter region to satisfy the upper bound on the dark matter relic abundance. 
In Sec.~4, we employ the results by the ATLAS and the CMS collaborations at the LHC Run-2 
  on search for the $Z^\prime$ boson resonance, and constrain the model parameter region.   
We find that the two parameter regions are complementary to each other, and  
  lead to a narrow allowed window in the ($m_{Z^\prime}, \alpha_{BL}$)-plane.  
The last section is devoted to conclusions.

\section{The minimal $B-L$ model with a dark matter candidate}
\begin{table}[t]
\begin{center}
\begin{tabular}{c|ccc|c|c}
      &  SU(3)$_c$  & SU(2)$_L$ & U(1)$_Y$ & U(1)$_{B-L}$  & $ Z_2 $\\ 
\hline
$q^{i}_{L}$ & {\bf 3 }    &  {\bf 2}         & $ 1/6$       & $1/3$    & $+$\\
$u^{i}_{R}$ & {\bf 3 }    &  {\bf 1}         & $ 2/3$       & $1/3$   & $+$\\
$d^{i}_{R}$ & {\bf 3 }    &  {\bf 1}         & $-1/3$       & $1/3$   &$+$\\
\hline
$\ell^{i}_{L}$ & {\bf 1 }    &  {\bf 2}         & $-\frac{1}{2}$       & $-1$  & $+$ \\
$e^{i}_{R}$    & {\bf 1 }    &  {\bf 1}         & $-1$                   & $-1$   & $+$ \\
\hline
$H$            & {\bf 1 }    &  {\bf 2}         & $-\frac{1}{2}$       & $0$  & $+$ \\  
\hline
$N^{j}_{R}$    & {\bf 1 }    &  {\bf 1}         &$0$                    & $-1$   & $+$ \\
$N_{R}$         & {\bf 1 }    &  {\bf 1}         &$0$                    & $-1$   & $-$ \\
$\Phi$            & {\bf 1 }       &  {\bf 1}       &$ 0$                  & $+2$  & $+$ \\ 
\hline
\end{tabular}
\end{center}
\caption{
The particle content of the minimal U(1)$_{B-L}$ extended SM with $Z_2$ parity. 
In addition to the SM particle content, the three right-handed neutrinos $N_R^j$ ($j=1,2$) and $N_R$ 
  and a complex scalar $\Phi$ are introduced. 
The $Z_2$ parity is also introduced, under which the right-handed neutrino $N_R$ is odd, 
  while the other fields are even.  
}
\label{table1}
\end{table}

We first define our model by the particle content listed on Table~1. 
The global $B-L$ symmetry in the SM is gauged, and the three right-handed neutrinos 
  ($N_R^1$, $N_R^2$  and $N_R$) and a $B-L$ Higgs field ($\Phi$) are introduced.  
The introduction of the $Z_2$ parity is crucial to incorporate a dark matter candidate in the model. 
Under this parity, the right-handed neutrino $N_R$ is assigned to be odd, 
  while the other fields are even. 
The conservation of the $Z_2$ parity ensures the stability of the $Z_2$-odd $N_R$, 
  and therefore, this right-handed neutrino is a unique dark matter candidate in the model~\cite{OS_DM}.

The Yukawa sector of the SM is extended to have 
\bea
\mathcal{L}_{Yukawa} \supset  - \sum_{i=1}^{3} \sum_{j=1}^{2} Y^{ij}_{D} \overline{\ell^i_{L}} H N_R^j 
          -\frac{1}{2} \sum_{k=1}^{2} Y^k_N \Phi \overline{N_R^{k~C}} N_R^k 
          -\frac{1}{2}  Y_N \Phi \overline{N_R^{~C}} N_R 
  + {\rm h.c.} ,
\label{Lag1} 
\eea
where the first term is the neutrino Dirac Yukawa coupling, and the second and third terms are the Majorana Yukawa couplings. 
Without loss of generality, the Majorana Yukawa couplings are already diagonalized in our basis.  
Note that because of the $Z_2$ parity only the two generation right-handed neutrinos are involved 
  in the neutrino Dirac Yukawa coupling. 
Once the $B-L$ Higgs field $\Phi$ develops the vacuum expectation value (VEV),  
  the $B-L$ gauge symmetry is broken and the Majorana mass terms for the right-handed neutrinos are generated. 
The seesaw mechanism~\cite{seesaw} is automatically implemented in the model 
  after the electroweak symmetry breaking.  
Because of the $Z_2$ parity, only two generation right-handed neutrinos are relevant to 
  the seesaw mechanism, and this so-called minimal seesaw~\cite{Minimal-Seesaw} possesses a number of  
  free parameters $Y_D^{ij}$ and $Y_N^k$ enough to reproduce the neutrino oscillation data 
  with a prediction of one massless eigenstate.\footnote{
When we consider leptogenesis scenario~\cite{FY}  in our model, 
  only two right-handed neutrinos are involved. 
See, for example,  \cite{IOO-LG} for detailed analysis of leptogenesis 
  with two right-handed TeV scale neutrinos.  
The model can successfully generate a sufficient amount of baryon asymmetry in the universe. 
}

The renormalizable scalar potential for the SM Higgs and the $B-L$ Higgs fields are generally given by 
\bea  
V = m_H^2 (H^{\dagger}H) + m_\Phi^2 (\Phi^{\dagger} \Phi) 
+ \lambda_H (H^{\dagger}H)^2 + \lambda_{\Phi} (\Phi^{\dagger} \Phi)^2 + \lambda_{H\Phi} (H^{\dagger}H)(\Phi^{\dagger} \Phi). 
\label{Higgs_Potential }
\eea
The parameters in the Higgs potential are suitably chosen for the Higgs fields to develop their VEVs as 
\bea
  \langle H \rangle =  \left(  \begin{array}{c}  
    \frac{v}{\sqrt{2}} \\
    0 \end{array}
\right),  \;  \;  \; 
\langle \Phi \rangle =  \frac{v_{BL}}{\sqrt{2}}. 
\eea
Associated with the $B-L$ symmetry breaking, the Majorana neutrinos $N_R^j$ $(j=1,2)$, 
  the dark matter particle $N_R$ and the $B-L$ gauge boson acquire their masses as 
\bea 
  m_N^j=\frac{Y_N^j}{\sqrt{2}} v_{BL},  \; \; 
  m_{DM}=\frac{Y_N}{\sqrt{2}} v_{BL},  \; \; 
  m_{Z^\prime} = 2 g_{BL} v_{BL}, 
\eea  
where $g_{BL}$ is the U(1)$_{B-L}$ gauge coupling.

The dark matter particle can communicate with the SM particles in two ways. 
One is through the Higgs bosons. 
In the Higgs potential of Eq.~(\ref{Higgs_Potential }), the SM Higgs boson and the $B-L$ Higgs boson 
  mix with each other in the mass eigenstates, and this Higgs boson mass eigenstates mediate  
  the interactions between the dark matter particle and the SM particles. 
Dark matter physics with the interactions mediated by the Higgs bosons have been investigated 
  in \cite{OS_DM, OO_DM}.  
The analysis involves 4 free parameters: Yukawa coupling $Y_N$ and  
  3 free parameters from the Higgs potential after two conditions of $v=246$ GeV 
  and the SM-like Higgs boson mass fixed to be 125 GeV are taken into account. 
The other way for the dark matter particle to communicate with the SM particles 
  is through the $B-L$ gauge interaction with the $Z^\prime_{BL}$ boson.  
In this case, only three free parameters ($g_{BL}$, $m_{Z^\prime}$ and $m_{DM}$) are involved 
   in dark matter physics analysis. 
As we have stated in the previous section, we concentrate on dark matter physics mediated 
   by the $Z^\prime_{BL}$ boson. 
When $|\lambda_{H \Phi}| \ll 1$, the Higgs bosons mediated interactions are negligibly small, 
  and the dark matter particle communicates with the SM particles only through the $Z^\prime_{BL}$ boson. 
For example, this situation is realized in supersymmetric extension of our model~\cite{MSUSYBL}, 
  where $\lambda_{H \Phi}$ is forbidden by supersymmetry in the Higgs superpotential 
  at the renormalizable level. 
When squarks and sleptons are all heavier than the dark matter particles, 
  there is no essential difference in dark matter phenomenology between non-supersymmetric case and 
  supersymmetric case (see Ref.~\cite{MSUSYBL}).  
For a limited parameter choice, the $Z^\prime_{BL}$ portal dark matter scenario has been investigated 
  in \cite{OO_DM, MSUSYBL}.

\section{Cosmological constraint on $Z^\prime_{BL}$ portal dark matter}
The dark matter relic abundance is measured at the 68\% limit as \cite{Planck2015}  
\bea 
   \Omega_{DM} h^2 = 0.1198\pm 0.0015.  
\eea 
In this section, we evaluate the relic abundance of the dark matter $N_R$ 
  and identify an allowed parameter region that satisfies the upper bound 
  on the dark matter relic density of $\Omega_{DM} h^2 \leq 0.1213$.  
The dark matter relic abundance is evaluated by integrating the Boltzmann equation given by 
\bea 
  \frac{dY}{dx}
  = - \frac{s(m_{DM}) \langle \sigma v \rangle}{x^2 H(m_{DM})} \left( Y^2-Y_{EQ}^2 \right), 
\label{Boltmann}
\eea  
where temperature of the universe is normalized by the mass of the right-handed neutrino $x=m_{DM}/T$, 
   $H(m_{DM})$ is the Hubble parameter at $T=m_{DM}$, 
   $Y$ is the yield (the ratio of the dark matter number density to the entropy density $s$) of 
  the dark matter particle, $Y_{EQ}$ is the yield of the dark matter particle in thermal equilibrium, 
  and $\langle \sigma v \rangle$ is the thermal average of the dark matter annihilation cross section times relative velocity. 
Explicit formulas of the quantities involved in the Boltzmann equation are as follows: 
\bea 
s(m_{DM}) &=& \frac{2  \pi^2}{45} \; g_\star \; m_{DM}^3 ,  \nonumber \\
H(m_{DM}) &=&  \sqrt{\frac{4 \pi^3}{45} g_\star} \; \frac{m_{DM}^2}{M_{Pl}},  \nonumber \\ 
s Y_{EQ}&=& \frac{g_{DM}}{2 \pi^2} \frac{m_{DM}^3}{x} K_2(x),   
\eea
where $M_{Pl}=1.22 \times 10^{19}$  GeV is the Planck mass, 
   $g_{DM}=2$ is the number of degrees of freedom for the dark matter particle, 
   $g_\star$ is the effective total number of degrees of freedom for particles in thermal equilibrium 
   (in the following analysis, we use $g_\star=106.75$ for the SM particles),  
   and $K_2$ is the modified Bessel function of the second kind.   
In our $Z^\prime_{BL}$ portal dark matter scenario, a pair of dark matter annihilates into the SM particles 
   dominantly through the $Z^\prime_{BL}$ exchange in the $s$-channel.  
The thermal average of the annihilation cross section is given by 
\bea 
\langle \sigma v \rangle = \left(s Y_{EQ} \right)^{-2} g_{DM}^2 
  \frac{m_{DM}}{64 \pi^4 x} 
  \int_{4 m_{DM}^2}^\infty  ds \; \hat{\sigma}(s) \sqrt{s} K_1 \left(\frac{x \sqrt{s}}{m_{DM}}\right) , 
\label{ThAvgSigma}
\eea
where the reduced cross section is defined as $\hat{\sigma}(s)=2 (s- 4 m_{DM}^2) \sigma(s)$ 
   with the total annihilation cross section $\sigma(s)$, and $K_1$ is the modified Bessel function of the first kind. 
The total cross section of the dark matter annihilation process $NN \to Z^\prime_{BL} \to f {\bar f}$ ($f$ denotes the SM fermions) 
   is calculated as 
\bea 
 \sigma(s)=\frac{\pi}{4} \alpha_{BL}^2  \frac{\sqrt{s (s-4 m_{DM}^2)}}
  {(s-m_{Z^\prime}^2)^2+m_{Z^\prime}^2 \Gamma_{Z^\prime}^2} 
  \left[ \frac{148}{9} + \frac{4}{3} \beta_t \left( 1- \frac{1}{3} \beta_t^2 \right)
  \right] 
\label{DMSigma}
\eea 
with $\beta_t(s)=\sqrt{1-4 m_t^2/s}$, top quark mass of $m_t=173.34$ GeV 
  and the total decay width of $Z^\prime_{BL}$ boson given by 
\bea
\Gamma_{Z'} = 
 \frac{\alpha_{BL}}{6} m_{Z^\prime} 
 \left[ \frac{37}{3} + \frac{1}{3} \beta_t(m_{Z^\prime}^2) \left( 3- \beta_t(m_{Z^\prime}^2)^2 \right)
 + \left( 1-\frac{4 m_{DM}^2}{m_{Z^\prime}^2} \right)^{\frac{3}{2}} 
 \theta \left( \frac{m_{Z^\prime}^2}{m_{DM}^2} - 4 \right)  \right]. 
\label{width}
\eea
Here, we have neglected all SM fermion masses except for $m_t$, 
  and assumed $m_N^j > m_{Z^\prime}/2$, for simplicity.

\begin{figure}[t]
\begin{center}
{\includegraphics[scale=1.3]{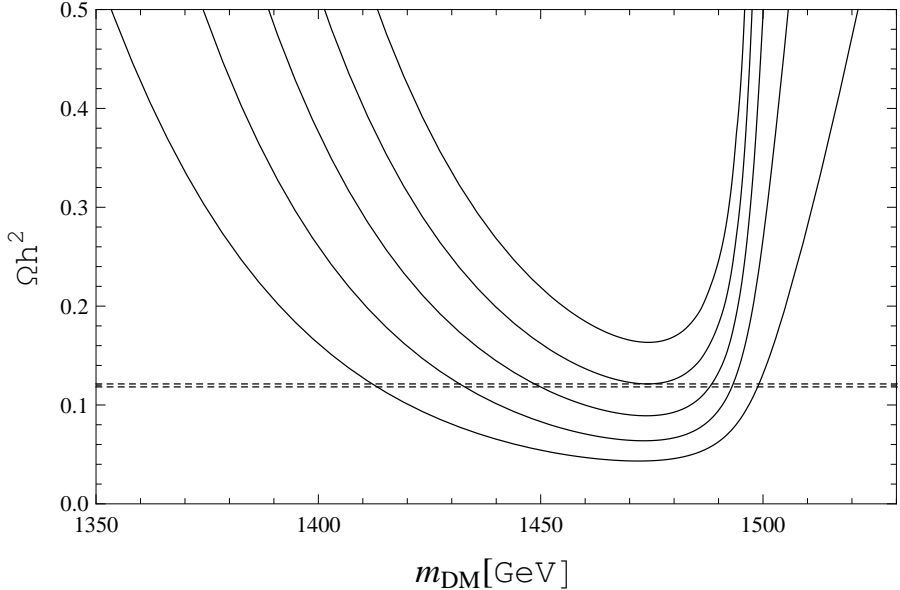}}
\caption{
The relic abundance of the $Z^\prime_{BL}$ portal right-hard neutrino dark matter 
  as a function of the dark matter mass ($m_{DM}$) for $m_{Z^\prime}=3$ TeV 
  and various values of the gauge coupling $\alpha_{BL}=0.001$, $0.0014$, $0.002$, $0.003$ and $0.005$ 
  (solid lines from top to bottom). 
The two horizontal lines denote the range of the observed dark matter relic density, 
  $0.1183 \leq \Omega_{DM} h^2 \leq 0.1213$.   
}
\label{fig1}
\end{center}
\end{figure}

Now we solve the Boltzmann equation numerically, and find the asymptotic value of the yield $Y(\infty)$. 
Then, the dark matter relic density is evaluated as 
\bea 
  \Omega_{DM} h^2 =\frac{m_{DM} s_0 Y(\infty)} {\rho_c/h^2}, 
\eea 
  where $s_0 = 2890$ cm$^{-3}$ is the entropy density of the present universe, 
  and $\rho_c/h^2 =1.05 \times 10^{-5}$ GeV/cm$^3$ is the critical density.
In our analysis, only three parameters, 
   namely $\alpha_{BL}=g_{BL}^2/(4 \pi)$, $m_{Z^\prime}$ and $m_{DM}$, are involved.  
For $m_{Z^\prime}=3$ TeV and various values of the gauge coupling $\alpha_{BL}$, 
  Fig.~\ref{fig1} shows the resultant dark matter relic abundance 
  as a function of the dark matter mass $m_{DM}$, 
  along with the observed bounds $0.1183 \leq \Omega_{DM} h^2 \leq 0.1213$~\cite{Planck2015} 
  (two horizontal dashed lines). 
The solid lines from top to bottom correspond to the results 
  for $\alpha_{BL}=0.001$, $0.0014$, $0.002$, $0.003$ and $0.005$, respectively. 
We can see that only if the dark matter mass is close to half of the $Z^\prime_{BL}$ boson mass, 
  the observed relic abundance can be reproduced. 
In other words, normal values of the dark matter annihilation cross section leads to over-abundance, 
  and it is necessary that an enhancement of the cross section through the $Z^\prime_{BL}$ boson resonance 
  in the $s$-channel annihilation process.

\begin{figure}[t]
\begin{center}
\includegraphics[width=0.45\textwidth,angle=0,scale=1.05]{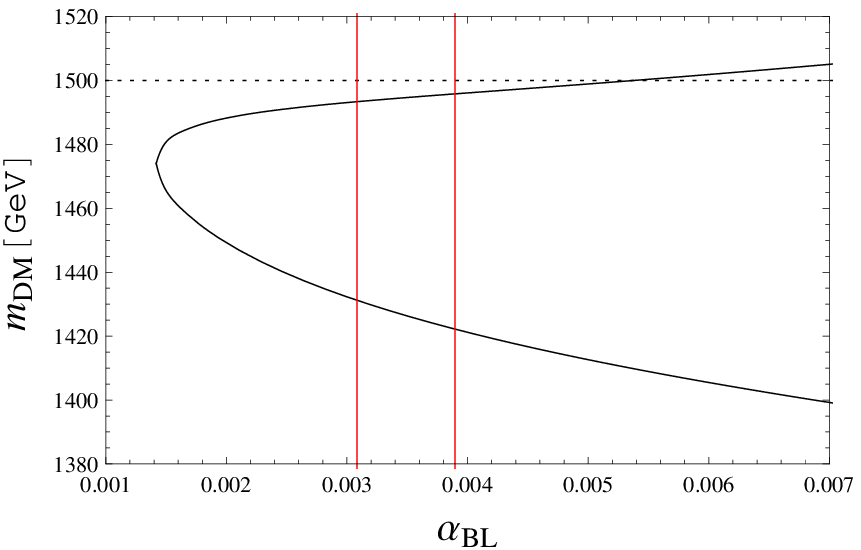} 
\hspace{0.1cm}
\includegraphics[width=0.45\textwidth,angle=0,scale=1.05]{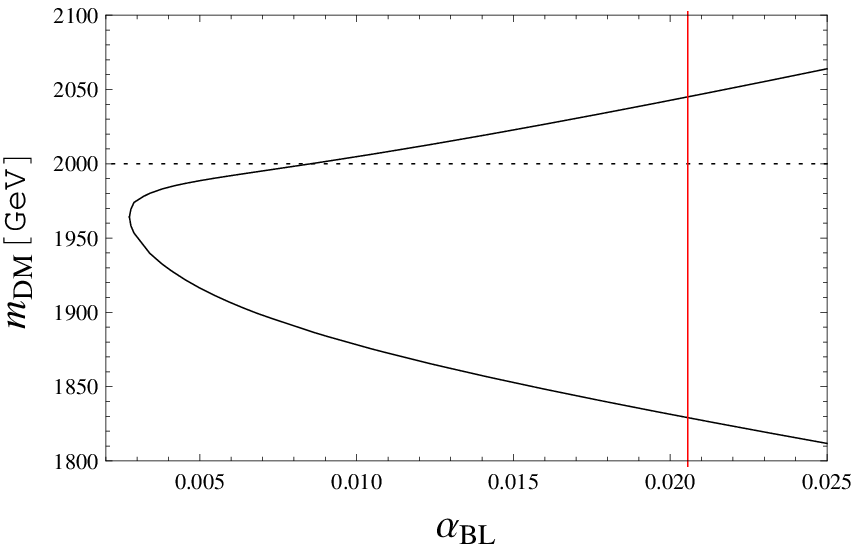}
\end{center}
\caption{
The dark matter mass as a function of $\alpha_{BL}$ for $m_{Z^\prime}=3$ TeV (left panel) 
  and $m_{Z^\prime}=4$ TeV (right panel). 
Along the solid (black) curve in each panel,  $\Omega_{DM} h^2 =0.1198$ is satisfied. 
The dotted lines correspond to $m_{DM}=m_{Z^\prime}/2$. 
The vertical solid lines (in red) denote the upper bound on $\alpha_{BL}$ 
  obtained from the recent LHC Run-2 results (see Figs.~\ref{fig4} and \ref{fig5}).    
In the left panel, the left vertical line represents the constraint from the ATLAS result~\cite{ATLAS13TeV},  
  while the right one is from the CMS result~\cite{CMS13TeV}.   
In the right panel, the vertical line represents the constraint from the ATLAS result~\cite{ATLAS13TeV}. 
}
\label{fig2}
\end{figure}

For a fixed $m_{DM}$ in the Fig.~\ref{fig1}, the resultant relic abundance becomes larger 
  as the gauge coupling $\alpha_{BL}$ is lowered. 
As a result, there is a lower bound on $\alpha_{BL}$ in order to satisfy the cosmological upper bound 
 on the dark matter relic abundance $\Omega_{DM} h^2 \leq 0.1213$. 
For a $\alpha_{BL}$ value larger than the lower bound, we can find two values of $m_{DM}$ 
  which result in the center value of the observed relic abundance $\Omega_{DM} h^2 =0.1198$. 
In Fig.~\ref{fig2}, we show the dark matter mass resulting $\Omega_{DM} h^2 =0.1198$ 
  as  a function of $\alpha_{BL}$.  
The left panel shows the result for $m_{Z^\prime}=3$ TeV, while the corresponding results for $m_{Z^\prime}=4$ TeV
  is shown in the right panel. 
As a reference, we also show the dotted lines corresponding to $m_{DM}=m_{Z^\prime}/2$. 
In Fig.~\ref{fig1}, we see that the minimum relic abundance is achieved 
   by a dark matter mass which is very close to, but smaller than $m_{Z^\prime}/2$.  
Although the annihilation cross section of Eq.~(\ref{DMSigma}) has a peak 
   at $\sqrt{s}=m_{Z^\prime}$, the thermal averaged cross section given in Eq.~(\ref{ThAvgSigma})  
   includes the integral of the product of the reduced cross section and 
   the modified Bessel function $K_1$. 
Our results indicate that for $m_{DM}$ taken to be slightly smaller than $m_{Z^\prime}/2$,  
   the thermal averaged cross section is larger than the one for $m_D=m_{Z^\prime}/2$.

\begin{figure}[t]
\begin{center}
{\includegraphics[scale=1.3]{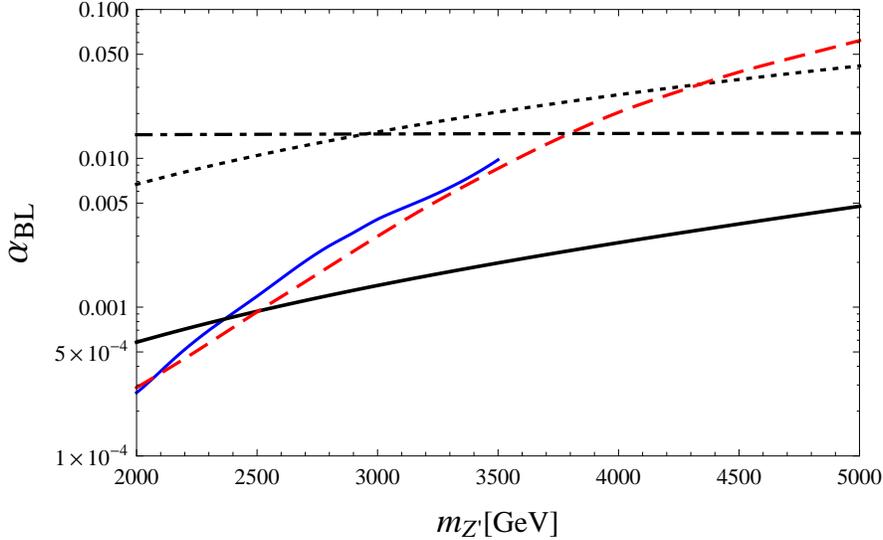}}
\caption{
Allowed parameter region for the $Z^\prime_{BL}$ portal dark matter scenario. 
The solid (black) line shows the lower bound on $\alpha_{BL}$ as a function of $m_{Z^\prime}$ 
  to satisfy the cosmological upper bound  on the dark matter relic abundance. 
The dashed line (in red) shows the upper bound on $\alpha_{BL}$ as a function of $m_{Z^\prime}$ 
  from the search results for $Z^\prime$ boson resonance by the ATLAS collaboration, 
  while the diagonal line (in blue) in the range of 2000 GeV$\leq m_{Z^\prime} \leq 3500$ GeV
  denotes the upper bound obtained from the result by the CMS collaboration.   
The LEP bound is depicted as the dotted line. 
The regions above these dashed, (blue) solid and dotted lines are excluded. 
We also show a theoretical upper bound on $\alpha_{BL}$ to avoid the Landau pole of 
  the running $B-L$ gauge coupling below the Planck mass $M_{Pl}$. 
}
\label{fig3}
\end{center}
\end{figure}

As mentioned above, for a fixed $Z^\prime_{BL}$ boson mass, we can find 
  a corresponding lower bound on the gauge coupling $\alpha_{BL}$ 
  in order for the resultant relic abundance not to exceed the cosmological 
  upper bound $\Omega_{DM} h^2 =0.1213$.  
Figure~\ref{fig3} depicts the lower bound of $\alpha_{BL}$ as a function of $m_{Z^\prime}$ (solid (black) line). 
Along this solid (black) line, we find that the dark matter mass is approximately given by 
  $m_{DM} \simeq 0.49~m_{Z^\prime}$. 
The dark matter relic abundance exceeds the cosmological upper bound  
  in the region below the solid (black) line. 
Along with the other constraints that will be obtained in the next section,  
  Fig.~\ref{fig3} is our main results in this paper.

\section{Interpretation of LHC Run-2 results}  
Very recently, the LHC Run-2 started its operation with a 13 TeV collider energy. 
Preliminary results from the ATLAS and the CMS collaborations have been reported~\cite{LHCreport}. 
The Run-2 results have provided constraints on new physics models, 
  some of which are more severe than those by the LHC  Run-1 results. 
The ATLAS and the CMS collaborations continue search for $Z^\prime$ boson resonance 
  with dilepton final states at the LHC Run-2, and have improved the upper limits of 
  the $Z^\prime$ boson production cross section from those in the LHC Run-1~\cite{ATLAS8TeV, CMS8TeV}. 
Employing the LHC Run-2 results, we will derive an upper bound on $\alpha_{BL}$ 
  as a function of $m_{Z^\prime}$. 
Since we have obtained in the previous section the lower bound on $\alpha_{BL}$  as a function of $m_{Z^\prime}$ 
  from the constraint on the dark matter relic abundance,  
  the LHC Run-2 results are complementary to the cosmological constraint.  
As a result, the parameter space of the $Z^\prime_{BL}$ portal dark matter scenario 
  is severally constrained once the two constraints are combined.

Let us calculate the cross section for the process $pp \to Z^\prime_{BL} +X \to \ell^{+} \ell^{-} +X$. 
The differential cross section with respect to the invariant mass $M_{\ell \ell}$ of the final state dilepton 
   is described as
\begin{eqnarray}
 \frac{d \sigma}{d M_{\ell \ell}}
 =  \sum_{a, b}
 \int^1_ \frac{M_{\ell \ell}^2}{E_{\rm CM}^2} dx
 \frac{2 M_{\ell \ell}}{x E_{\rm CM}^2}  
 f_a(x, Q^2) f_b \left( \frac{M_{\ell \ell}^2}{x E_{\rm CM}^2}, Q^2
 \right)  {\hat \sigma} (q \bar{q} \to Z^\prime_{BL} \to  \ell^+ \ell^-) ,
\label{CrossLHC}
\end{eqnarray}
where $f_a$ is the parton distribution function for a parton ``a'', 
  and $E_{\rm CM} =13$ TeV is the center-of-mass energy of the LHC Run-2.
In our numerical analysis, we employ CTEQ6L~\cite{CTEQ} for the parton distribution functions 
   with the factorization scale $Q= m_{Z^\prime}$. 
Here, the cross section for the colliding partons is given by 
\bea 
{\hat \sigma} = 
\frac{4 \pi}{81}  \alpha_{BL}^2  
\frac{M_{\ell \ell}^2}{(M_{\ell \ell}^2-m_{Z^\prime}^2)^2 + m_{Z^\prime}^2 \Gamma_{Z^\prime}^2}. 
\label{CrossLHC2}
\eea
By integrating the differential cross section over a range of $M_{\ell \ell}$ set by the ATLAS and the CMS analysis, 
  respectively, we obtain the cross section to be compared with the upper bounds 
  obtained by the ATLAS and the CMS collaborations.

\begin{figure}[t]
\begin{center}
\includegraphics[width=0.45\textwidth,angle=0,scale=1.06]{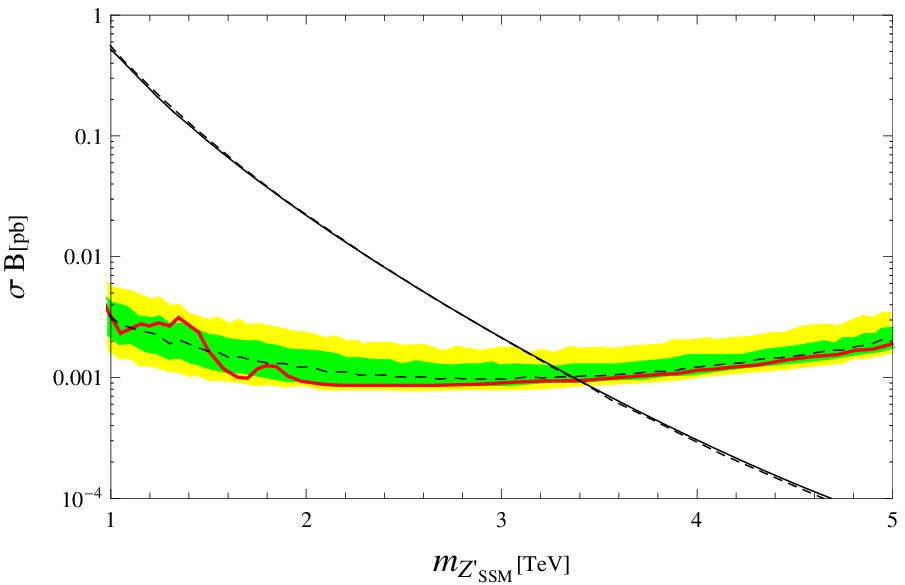} 
\hspace{0.1cm}
\includegraphics[width=0.45\textwidth,angle=0,scale=1.08]{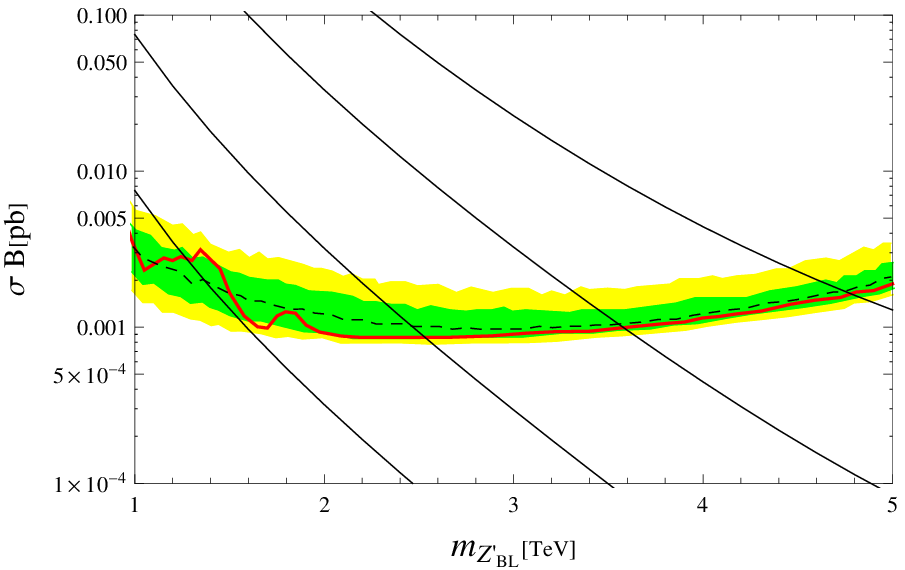}
\end{center}
\caption{
Left panel: the cross section as a function of the $Z^\prime_{SSM}$ mass (solid line) 
  with $k=1.31$, along with the ATLAS result in Ref.~\cite{ATLAS13TeV}
   from the combined dielectron and dimuon channels. 
Right panel: the cross sections calculated for various values of $\alpha_{BL}$ with $k=1.31$. 
The solid lines from left to right correspond to $\alpha_{BL}=0.0001$, $0.001$, $0.01$ and $0.05$, respectively. 
}
\label{fig4}
\end{figure}

In the analysis by the ATLAS and the CMS collaborations, 
   the so-called sequential SM $Z^\prime$ ($Z^\prime_{SSM}$) model~\cite{ZpSSM} 
   has been considered as a reference model.  
We first analyze the sequential $Z^\prime$ model to check a consistency of our analysis 
   with the one by the ATLAS collaboration.  
In the sequential $Z^\prime$ model, the $Z^\prime_{SSM}$ boson has exactly the same 
   couplings with quarks and leptons as the SM $Z$ boson. 
With the couplings, we calculate the cross section of the process $pp \to Z^\prime_{SSM}+X \to \ell^+ \ell^- +X$ 
   like Eq.~(\ref{CrossLHC}). 
By integrating the differential cross section in the region of 128 GeV$\leq M_{\ell \ell} \leq 6000$ GeV~\cite{ATLAS8TeV}, 
  we obtain the cross section of the dilepton production process as a function of $Z^\prime_{SSM}$ boson mass.\footnote{
Since the decay width of the $Z^\prime_{SSM}$ boson is narrow, the cross section is almost determined 
  by the integral in the vicinity of the resonance pole.} 
Our result is shown as a solid line in the left panel on Fig.~\ref{fig4}, 
  along with the plot presented by the ATLAS collaboration~\cite{ATLAS13TeV}.  
In the analysis in the ATLAS paper, the lower limit of the $Z^\prime_{SSM}$ boson mass is found to be $3.4$ TeV,  
   which is read from the intersection point of the theory prediction (diagonal dashed line) and  
   the experimental cross section bound (horizontal solid curve (in red)).  
In order to take into account the difference of the parton distribution functions used in the ATLAS and our analysis 
  and QCD corrections of the process, we have scaled our resultant cross section by a factor $k=1.31$, 
  with which we can obtain the same lower limit of the $Z^\prime_{SSM}$ boson mass as $3.4$ TeV.  
We can see that our result with the factor of $k=1.31$ is very consistent with the theoretical prediction 
  (diagonal dashed line) presented in Ref.~\cite{ATLAS13TeV}. 
This factor is used in our analysis of the $Z^\prime_{BL}$ production process. 
Now we calculate the cross section of the process  $pp \to Z^\prime_{BL}+X \to \ell^+ \ell^- +X$ 
   for various values of $\alpha_{BL}$,  
   and our results are shown in the right panel of Fig.~\ref{fig4}, along with the plot in Ref.~\cite{ATLAS13TeV}. 
The diagonal solid lines from left to right correspond to $\alpha_{BL}=0.0001$, $0.001$, $0.01$ and $0.05$, respectively. 
From the intersections of the horizontal curve and diagonal solid lines, we can read off a lower bound 
  on the $Z^\prime_{BL}$ boson mass for a fixed $\alpha_{BL}$ value.  
In this way, we have obtained the upper bound on $\alpha_{BL}$ as a function of the $Z^\prime_{BL}$ boson mass, 
  which is depicted in Fig.~\ref{fig3} (dashed (red) line).

\begin{figure}[t]
\begin{center}
\includegraphics[width=0.45\textwidth,angle=0,scale=1.06]{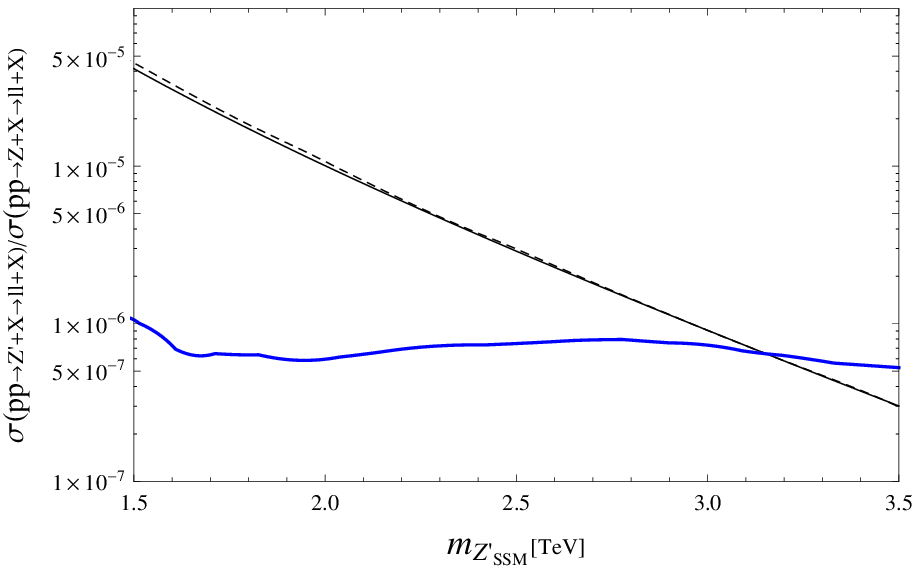} 
\hspace{0.1cm}
\includegraphics[width=0.45\textwidth,angle=0,scale=1.07]{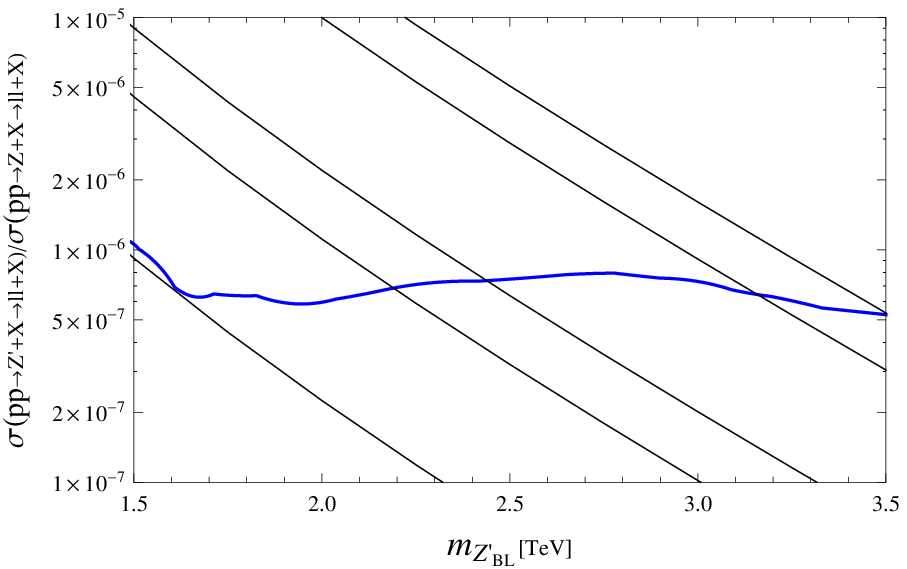}
\end{center}
\caption{
Left panel: the cross section ratio as a function of the $Z^\prime_{SSM}$ mass (solid line) 
  with $k=1.80$, along with the CMS result in Ref.~\cite{CMS13TeV}
   from the combined dielectron and dimuon channels. 
Right panel: the cross section ratios calculated for various values of $\alpha_{BL}$ with $k=1.80$. 
The solid lines from left to right correspond to 
  $\alpha_{BL}=0.0001$, $0.0005$, $0.001$, $0.005$ and $0.01$, respectively. 
}
\label{fig5}
\end{figure}

We apply the same strategy and compare our result for the $Z^\prime_{SSM}$ model 
   with the one by the CMS collaboration~\cite{CMS13TeV}.
According to the CMS analysis, we integrate the differential cross section in the range of 
  $0.97 \; m_{Z^\prime_{SSM}} \leq  M_{\ell \ell} \leq  1.03 \; m_{Z^\prime_{SSM}}$.  
In the CMS analysis, a limit has been set on the ratio of the $Z^\prime_{SSM}$ boson cross section 
  to the $Z/\gamma^*$ cross section in a mass window of 60 to 120 GeV, 
  which is predicted to be $1928$ pb.   
Our result is shown as a diagonal solid line in the left panel of Fig.~\ref{fig5}, 
  along with the plot presented in Ref.~\cite{CMS13TeV}.  
The analysis in this CMS paper leads to the lower limit of the $Z^\prime_{SSM}$ boson mass as $3.15$ TeV,  
   which is read from the intersection point of the theory prediction (diagonal dashed line) and  
   the experimental cross section bound (horizontal solid curve (in blue)).  
In order to obtain the same lower mass limit of $m_{Z^\prime_{SSM}} \geq 3.15$ TeV,  
   we have introduced a factor $k=1.80$.  
The left panel shows that our results are very consistent with the theoretical cross section  
   presented in Ref.~\cite{CMS13TeV}.

With the factor of $k=1.80$, we calculate the cross section of the process  $pp \to Z^\prime_{BL}+X \to \ell^+ \ell^- +X$ 
   for various values of $\alpha_{BL}$,  
   and our results are shown in the right panel of Fig.~\ref{fig5}, along with the plot in Ref.~\cite{CMS13TeV}. 
The diagonal solid lines from left to right correspond to 
  $\alpha_{BL}=0.0001$, $0.0005$, $0.001$, $0.005$ and $0.01$, respectively. 
From the intersections of the horizontal (blue) curve and diagonal solid lines, we can read off a lower bound 
  on the $Z^\prime_{BL}$ boson mass for a fixed $\alpha_{BL}$ value.  
In Fig.~\ref{fig3}, the diagonal solid (blue) line in the range of 2000 GeV$\leq m_{Z^\prime_{BL}} \leq 3500$ GeV  
   shows the upper bound on $\alpha_{BL}$ as a function of the $Z^\prime_{BL}$ boson mass. 
The ATLAS and the CMS bounds we have obtained are consistent with each other. 
The ATLAS bound is slightly more severe than the CMS bound, and applicable to a higher mass range 
   up to $m_{Z^\prime} =5000$ GeV.

In Fig.~\ref{fig3}, we also show the LEP bound as the dotted line 
  which is obtained from the search for effective 4-Fermi interactions 
  mediated by the $Z^\prime_{BL}$ boson~\cite{LEPbound1}. 
An updated limit with the final LEP 2 data~\cite{LEP2data} is found to be~\cite{LEPbound2} 
\bea
   \frac{m_{Z^\prime}}{g_{BL}} \geq 6.9 \; {\rm TeV}  
\eea 
at 95\% confidence level. 
We find that the ATLAS bound at the LHC Run-2 is more severe than the LEP bound 
  for $m_{Z^\prime} \lesssim 4.3$ TeV.  
In order to avoid the Landau pole of the running $B-L$ coupling $\alpha_{BL}(\mu)$ 
  below the Plank mass, $1/\alpha_{BL}(M_{Pl}) > 0$, we find  
\bea
  \alpha_{BL} < \frac{\pi}{6 \ln \left[ \frac{M_{Pl}}{m_{Z^\prime}} \right]}, 
\eea 
which is shown as the dashed-dotted line in Fig.~\ref{fig3}.   
Here, the gauge coupling $\alpha_{BL}$ used in our analysis for dark matter physics and LHC physics 
  is defined as the running gauge coupling $\alpha_{BL}(\mu)$ at $\mu=m_{Z^\prime}$, 
  and we have employed the renormalization group equation at the one-loop level 
  with $m_N^1=m_N^2=m_\Phi=m_{Z^\prime}$, for simplicity. 

\section{Conclusions}
We have considered the minimal gauged $B-L$ extension of the Standard Model,  
   which is free from all the gauge and gravitational anomalies and 
   automatically incorporates the neutrino mass and flavor mixing through the seesaw mechanism. 
We have extended this model by introducing a $Z_2$  parity, so that a dark matter candidate 
   is supplemented and identified as an $Z_2$-odd right-handed neutrino.   
No extension of the particle content from the one of the minimal $B-L$ model is needed.  
In this model, the dark matter particle communicates with the Standard Model particles 
  through the $B-L$ gauge boson ($Z^\prime_{BL}$ boson). 
Since the $B-L$ charges for all particles are fixed, physics of this ``$Z^\prime_{BL}$ portal'' dark matter scenario 
  is controlled by only three parameters, namely, the gauge coupling, the $Z^\prime_{BL}$ boson mass, 
  and the dark matter mass.  
Imposing the cosmological upper bound on the dark matter relic density, 
  we have found the lower bound on the $B-L$ gauge coupling as a function 
  of the $Z^\prime_{BL}$ boson mass.   
Search results for $Z^\prime$ boson resonance by the ATLAS and CMS collaborations at the LHC Run-2 
  provide the information that is complementary to the cosmological bound 
  on the ``$Z^\prime_{BL}$ portal'' dark matter scenario.  
We have interpreted the $Z^\prime$ boson resonance search results at the LHC Run-2, and obtained the upper bound 
  on the $B-L$ gauge coupling as a function of the $Z^\prime_{BL}$ boson mass.  
Similar upper bounds on the $B-L$ gauge coupling can be obtained through  
  results by the LEP experiment of search for effective 4-Fermi interactions mediated 
  by the $Z^\prime_{BL}$ boson and the requirement to maintain the running $B-L$ gauge coupling 
  in perturbative regime up to the Planck mass.  
Putting all together, our final result is shown in Fig.~\ref{fig3}. 
We have identified the allowed parameter region for the ``$Z^\prime_{BL}$ portal'' dark matter scenario, 
   which turns out to be narrow and leads to the lower bound on the $Z^\prime_{BL}$ boson mass 
   of $m_{Z^\prime} > 2.5$ TeV.

In the present model, the Standard Model fermions couple with the $Z^\prime_{BL}$ boson 
   through the vector current, while the dark matter particle has the axial current coupling 
   because of its Majorana nature. 
Hence, the elastic scattering cross section of the dark matter particle with nuclei
   vanishes in the non-relativistic limit, and the direct and indirect search for 
   the dark matter particle is not applicable to the present scenario.  
Our model can be easily extended to have more general U(1) gauge symmetry~\cite{generalU1},
  while keeping the same minimal particle content.  
In this case, the axial vector couplings between the Standard Model fermions 
  with the $Z^\prime$ gauge boson arise in general, and the dark matter particle 
  can scatter off nuclei.
In the context of the sequential $Z^\prime$ model as a reference, 
  the constraints from the direct and indirect dark matter search on the $Z^\prime$ portal dark matter scenario 
  have been investigated in Ref.~\cite{Zp-protal4}.   
Several representative $Z^\prime$ portal dark matter models have been examined 
  to account for the Galactic Center gamma-ray excess~\cite{Zp-protal5}.     
It is worth investigating this direction with the general U(1) extension of our scenario 
  with the right-handed neutrino dark matter~\cite{OO2}.

\section*{Acknowledgments}
S.O. would like to thank Ryusuke Endo for encouraging her to carry out the dissertation research.  
She also wishes to thank Shinsuke Kawai for his valuable advices. 
S.O. would like to thank the Department of Physics and Astronomy at the University of Alabama
  for hospitality during her visit for the completion of this work. 
We would like to record our gratitude to Andy Okada for his encouragements.  
The work of N.O. is supported in part by the United States Department of Energy grant (DE-SC0013680).


\end{document}